\documentclass[a4paper,12pt,reqno,superscriptaddress,nofootinbib]{revtex4}
\usepackage[centertags]{amsmath}
\usepackage{amsfonts}
\usepackage{amssymb}
\usepackage{amsthm}
\usepackage{newlfont}
\usepackage{stmaryrd}
\usepackage{mathrsfs}
\usepackage{mathtools}
\usepackage{euscript}
\usepackage{graphicx}
\usepackage{enumerate}
\usepackage{todonotes}

\usepackage{color}

\usepackage{tikz}
\usepackage{pgf}
\usetikzlibrary{positioning,fit,calc}
\usetikzlibrary{arrows,automata}
\usepackage{wrapfig}
\usepackage{subfigure}
\usepackage{amscd}
\usepackage{hyperref}

\usepackage{changes}

\theoremstyle{plain}

\theoremstyle{definition}

\theoremstyle{remark}

\let\ve=\varepsilon

\newcommand{\be}{\begin{equation}}
\newcommand{\en}{\end{equation}}

\newcommand{\opunit}{\text{1}\kern-0.22em\text{l}}

\newcommand{\id}{\textrm{d}}

\DeclareMathAlphabet{\mathpzc}{OT1}{pzc}{m}{it}

\let\oldsqrt\sqrt
\def\sqrt{\mathpalette\DHLhksqrt}
\def\DHLhksqrt#1#2{%
	\setbox0=\hbox{$#1\oldsqrt{#2\,}$}\dimen0=\ht0
	\advance\dimen0-0.2\ht0
	\setbox2=\hbox{\vrule height\ht0 depth -\dimen0}%
	{\box0\lower0.4pt\box2}}

\let\ve=\varepsilon

\let\be=\beta

\DeclareMathAlphabet{\mathpzc}{OT1}{pzc}{m}{it}

\def\bea{\begin{eqnarray}}
\def\eea{\end{eqnarray}}
\def\ba{\begin{array}}
	\def\ea{\end{array}}

\usepackage{changes}

\begin{document}

\title{Statistical mechanical foundation of Weber-Fechner laws}

\author{Christian Maes\\ {\it Instituut voor Theoretische Fysica, KU Leuven}}

\email{christian.maes@kuleuven.be}


\begin{abstract}
Even though the phenomenological relations between perception and stimulus have been firmly established, a theoretical argument for Weber's and Fechner's law in terms of relevant models or from statistical physics is largely missing.  We present such a discussion in terms of response theory for nonequilibrium systems, where the induced displacement or current, which stands for the perceived stimulus, crucially depends on the change in time-symmetric reactivities.  Stationary nonequilibria may indeed generate extra currents by changing the dynamical activity.  The argument finishes by understanding how the extra dynamical activity logarithmically encodes the actual stimulus.

\end{abstract}
\maketitle


\section{Introduction}
Weber's law (1834) and Fechner's extension (1860) summarize the basic phenomenology of perception.  They belong to the field of psychophysics and neuronal physiology more generally, which has been of central interest to physics from the time of Helmholtz to recent revivals in what is now called the field of neurophysics.\\
Weber's law speaks about the accuracy of discrimination.  Whether by touch, sound, light or smell, whenever we compare two sensations, there appears a limit to the accuracy of our discrimination and that keenness  varies, at least in many cases, with the magnitude of the agitation. For example, if for a load we are merely able to distinguish a mass of 100g from one of 108 gram, we would typically need a difference of 80g to notice the unequity of two masses of about 1kg.  That is, the just-noticeable-difference between the magnitudes of two stimuli increases proportional to the magnitude.\\  Fechner continued from there to state that the perceived stimulus is logarithmic in the actual stimulus:  $I_p = K \log I/I_0$ for a constant $K$, where $I$ is the actual intensity of the stimulus when $I>I_0$ and $I_p$ is the perceived one; $I_0$ can be identified with the treshold for perception. E.g., $I$ could be the luminous intensity of a source and $I_p$ would be the apparent brightness for eye light-intensity response.\\
It is no surprise that the Weber law about the accuracy of discrimination  is a consequence of the Fechner law $I_p = K \log (I/I_0)$ for the relation between perceived and actual stimulus intensities.  We only need to state that a difference in perception is possible at $\Delta I_p \sim 1$ (arbitrary units) to find that at that threshold $K \Delta I \sim I$ implying we need a difference $\Delta I$ in actual stimulus growing in proportion with the absolute intensity $I$. Therefore, no extra explanation of Weber's law is needed beyond that interpretation of Fechner's relation. \\
Those laws constitute a regularity in sensation that has been confirmed in a great variety of experiments, {\it auditu et tactu} as written in the title of \cite{webr}, over a wide range of stimulus strengths.  It has been observed with many animal species as well, and in much greater variety of sensations than originally discussed in the works of Weber and Fechner \cite{webr,fech}.  There are a number of generalizations and corrections, \cite{stevr,rev}, depending on the type of stimulus and on their range but in all, Weber-Fechner laws stand out as widely established phenomenology of psychophysics.\\

Over the last two centuries, from the sensory and nerve physiology pioneered by von Helmholtz to today's  neuroscience, imaging data and knowledge about perception and sensation have immensily increased without clearly revealing however the physics behind that ``foundation stone of experimental psychology,'' \cite{titch}.
Sure enough, analogies have been made and equivalences exposed between Weber-Fechner behavior and certain chemophysical or neuronal properties.  For example in \cite{cope}, the Fechner law is  associated to charge transport over solid-solid or solid-liquid interfaces, assuming the response behavior of the receptor follows	the Elovich equation  and that the generation of response is simply and directly proportional to stimulus intensity.  As a more recent example, in \cite{neuro}  Weber's law is tied to the dependence of reaction times under stimuli; see also \cite{rat} for the underlying decision model. Nevertheless, there is no uniquely or broadly accepted explanation of the Weber-Fechner laws from more elementary physics principles \cite{rev}.\\
There are probably a number of reasons for this lack of foundation, not only that the laws are approximate anyhow.  First, after the pioneering work in the 19th century, the phenomenology has been known and discussed most of all in the psychology literature and in the context of human preception, which obviously does not simplify matters for physics.    The Weber-Fechner laws have a vagueness or subjectivity in their `human' formulation, using such terms as ``accuracy of discrimination,'' ``threshold of sensation'' and  ``perceived perception,'' which makes them less accessible for a mechanistic explanation.  Objective `material' facts seem to be missing from the laws until the advent of modern neurophysiology.   There is however also a second class of reasons why the Weber-Fechner laws have remained elusive, and that has to do with the status of response theory in physics.  For the most part, in teaching and in applications, that has been restricted to linear response and to equilibrium systems.  The so-called fluctuation-dissipation theorem, the Einstein relation and Kubo theory of linear response traditionally deal with the close-to-equilibrium regime and pre-suppose a linear regime where force and displacement are linearly related \cite{ku,kubo,chan}.  Those criteria are of course not met for human or animal perception. Stimuli need not be small and the whole sensory equipment is subject to important active processes where fluctuation-dissipation relations are violated; see e.g.~\cite{pao,frog}. \\

Also for the line of arguments in the present paper it is essential to mention that perception can be split into two basic processes, an initial stage where the stimulus is coded in neuronal activity and a final stage where brain activity is coupled to cognitive and muscular processes creating the perceived stimulus.  Our ambition is not to model that full and complicated sequence of transmissions and conversions and we remain far from neurobiological details.  Our main innovation regards the final stages of the perception process where we interpret perceived stimulus as current or displacement induced by a collection of activated components.    In the main part we derive from response theory how excess activity determines the induced forces and currents.  Response  for active and nonequilibrium systems is a subject of much recent interest and we use the ideas in e.g. \cite{resp,fdr,urna,fdr1}.   For the initial stage of perception, only in Section \ref{webr} do we add to model the conversion of stimulus into neuronal activity, which is where the logarithm happens.   We argue for the naturalness of such encoding of the actual stimulus as excess reactivity.\\
The present paper is thus adding the following ingredients to arrive at a statistical foundation of Weber-Fechner laws: \\
 1)   Interpreting perception or the perceived stimulus as  displacement (e.g. in generating a shift in positions or velocities) we get rid of the subjective formulation and we make the Weber-Fechner laws a subject of response theory.  The actual stimulus is assumed to perturb (only) the time-symmetric reactivity in a driven Markov jump process modeling neuronal activity;\\
  2) Applying nonequilibrium response theory, we show that the induced excess in dynamical activity governs that displacement.  Therefore, from arguing logarithmic dependence in the coding of a stimulus in neuronal activity, we arrive at the Weber-Fechner laws.\\

  In Section \ref{nonr} we connect the Weber-Fechner phenomenology with response theory. It is the time-symmetric dynamical activity which is changed by the stimulus and which determines a corresponding displacement or perception. We first present in Section \ref{acti} two simple models to illustrate that main idea.  In Section \ref{webr} we end with arguments for the log-dependence of reaction rates on stimuli. Note also that the logarithmic response does not always get validated by the empirical facts; there are exceptions and limitations.  We briefly discuss in Section \ref{lims} how and where those may arise, such as in the power-law response summarized in Stevens' law \cite{stevr}.

\section{Perceived stimulus from excess activity}\label{acti}

Two simple examples can illustrate how excess in dynamical activity (visible in escape rate and reactivity) contributes to and even determines the current when fixing the out-of-equilibrium condition in terms of driving. In the first example, we consider driven passive particles for which the current gets amplified by increasing the dynamical activity.   The example has no relation with neuronal physiology, but illustrates the principle. In the second example, the particles are active and no net current needs to be present, until coupled with firing bits.  That example is already providing a toy-modeling of later stages in perception.  In all events, in the present section we are only concerned with the issue how changes in dynamical activity generate (extra) displacement.

\subsection{Passive model}\label{spins}
A dilute collection of charges are driven through a narrow tube.  The solvent is a viscous fluid in equilibrium at temperature $T$. There is a constant force ${\cal E}>0$ on the charges, say to the right.  Imagining that the tube is spatially periodic in one dimension with cells of size $L$, we take the model of a biased continuous-time random walk;  integers $x$ correspond to the cells in the tube.  The transition rates to hop to the right-neighboring, respectively to the left-neighboring cell, are 
\[
k(x,x+1) = p,\qquad k(x,x-1) = q
\]
where we require that $p/q = \exp[{\cal E}L/k_BT]$, expressing that the ratio of forward to  backward rates is given by the entropy flux to the environment in units of the Boltzmann constant $k_B$ (condition of local detailed balance; see \cite{resp}).   
 Writing ${\cal E}L/k_BT = \ve$, we thus have
\begin{equation}\label{rwr}
k(x,x+1) = \xi(\ve)\,\frac{1}{1+e^{-\ve}},\qquad k(x,x-1) = \xi(\ve)\,\frac 1{1+ e^{\ve}}
\end{equation}
where the escape rate $\xi(\ve) := p +q$ may well depend on the driving $\cal E$.  We fix a large $\ve$; that is the reference condition of the passive nonequilibrium system.  At that moment, the expected current, the net displacement of particles from cell to cell,  equals
\begin{equation}\label{ve}
\langle v\rangle_\ve= L\,( p - q) = L\,\xi(\ve)\,\frac{\sinh \ve}{1+ \cosh\ve} 
\end{equation}
\begin{figure}[!h]
	\centering
	\includegraphics[width=0.65\textwidth]{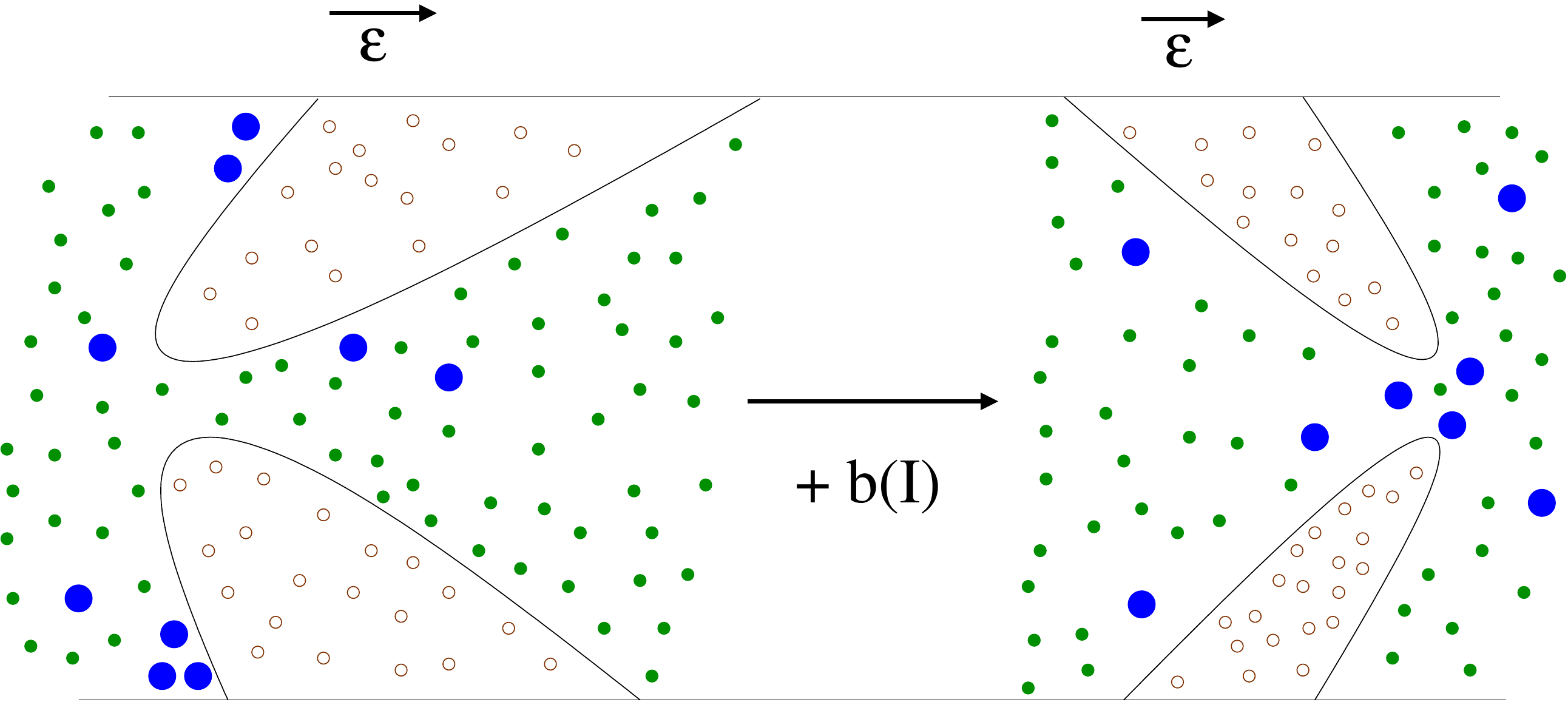}
	\caption{Larger (blue) spheres are driven by an external field $\ve$ in a tube with an irregular interior, dissipating into the viscous liquid represented by smaller (green) spheres.  There are obstacles drawn as blobs sticking out of the wall.  Left, before the extra activity $b(I)$ has been added: the escape rate $\xi(\ve)$ may get very low in high driving $\ve$.  Right, the extra activity $b(I)$ has increased the escape rate and hence the flux. Figure courtesy of Tirthankar Banerjee.}
	\label{tube}
\end{figure} 
 Note that it is the escape rate $\xi(\ve)$ that for large $\ve$ decides the current behavior \eqref{ve}.  It may very well be that $\xi(\ve) \downarrow 0$ for large $\ve$, depending on trapping or caging mechanism such as in the left picture of Fig.~\ref{tube}.  The standard fluctuation--dissipation relation is violated for large $\ve$.  In contrast, to linear order in $\ve$ the dependence $\xi(\ve)$ is irrelevant except for $\xi(0)\neq 0$: $\langle v\rangle_\ve= L\,\xi(\ve=0)\,\ve/2$.    See also \cite{negheatcap,chemfalasco}. \\
 Next, a stimulus of some sort is applied with intensity $I$, shaking up each cell in such a way, we assume, that the escape rate changes according to
\begin{equation}\label{stim}
\xi(\ve) \;\stackrel{\text{stimulus}}{\longrightarrow} \;\xi(\ve,I) = \xi(\ve) + b(I)
\end{equation}
for  excess parameter $b= b(I)$ growing with $I$.  We have not specified the mechanism, we only assumed that the stimulus modified  the escape rate $\xi(\ve)$, suggesting that on average the dynamical activity of the charges has increased.  See Fig.~\ref{tube}  for a cartoon.

We interpret the change in the current as the perceived stimulus:
\begin{equation}\label{vel}
I_p = \langle v\rangle_\ve^I - \langle v\rangle_\ve = L\,\,b(I) - L\,\xi(\ve=\infty)
\end{equation}
for large driving $\ve\uparrow \infty$.  The point here is that the extra reactivity $b(I)$ decides the displacement in current, $I_p \propto b(I)$ when $\xi(\ve=\infty) \simeq 0$.   Looking ahead to Section \ref{webr}: when the excess in escape rate is logarithmic in the stimulus strength $I$, $b(I)\sim \log I$, from \eqref{vel} we get Weber-Fechner phenomenology $I_p\sim \log I + \text{constant}$.

\subsection{Active model}\label{spinsa}
\begin{figure}[!h]
	\centering
	\includegraphics[width=0.65\textwidth]{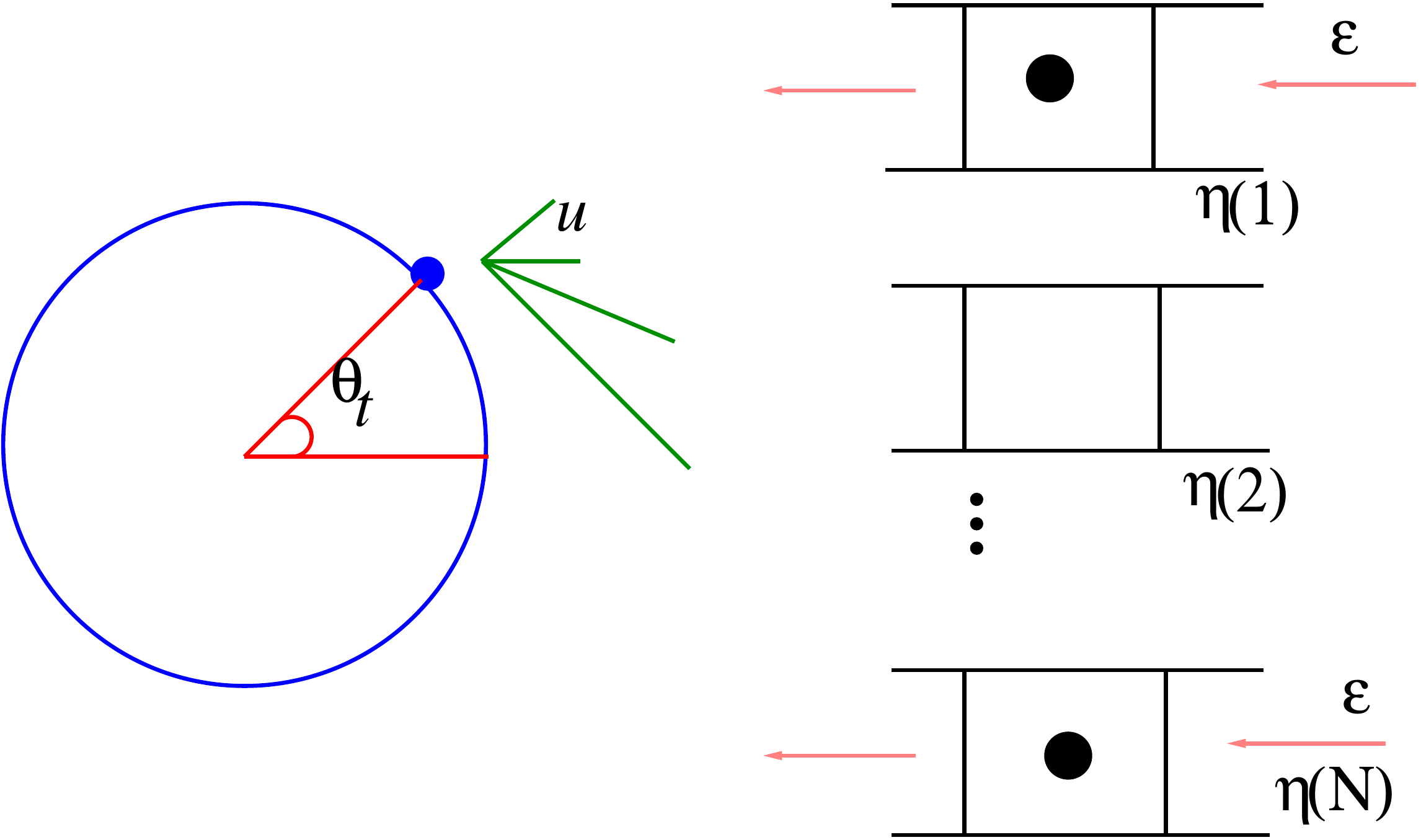}
	\caption{Firing bits $\eta_t(i)$ coupled to angle variable $\theta_t$ via potential $u$.  The nonequilibrium is fixed by parameter $\ve$, allowing the extra activity $a\rightarrow a_\ell =a(1 + \lambda\,b\, \cos\theta)$ to create angular flux.  Figure courtesy of Tirthankar Banerjee.}
	\label{fig2}
\end{figure}

Imagine next  motion on the circle $S^1$ characterized by an angle $\theta_t \in [0,2\pi]$ and coupled to $N$ bits $\eta_t(i) = 0, 1$.  The angle-coordinate follows the overdamped equation
\begin{equation}\label{act}
\dot \theta_t = -U'(\theta_t) -u'(\theta_t)\,\left(\eta_t(1) +\ldots+\eta_t(N)\right)
\end{equation}
for energy functions $U$ and $u$, where we ignore thermal noise and put the mobility to one (per second).  The energies
are defined in units of $k_BT$ for fixed environment temperature $T$.  The idea is that the angular velocity represents the perceived stimulus $I_p$ and the bit-dynamics is a spacetime conversion of the actual stimulus $I$.   Whenever some $\eta_t(i)=1$ an extra push is given to the angle.  Each bit $\eta_t(i)$ independently flips $0 \stackrel{\ell,r}{\longleftrightarrow} 1$ over two possible channels, left ($\ell$) and right ($r$),  with transition rates that depend on the angle,
\begin{align}
k_{\ell}(0,1) &= a_{\ell}(\theta)\,e^{-\frac{1}{2}[u(\theta) +\ve]},\quad k_{r}(0,1) = a\,e^{-\frac{1}{2}[u(\theta) - \ve]}\label{act1}
\\
k_{\ell}(1,0) &= a_{\ell}(\theta)\,e^{\frac{1}{2}[u(\theta) +\ve]},\quad k_{r}(1,0) = a\,e^{\frac{1}{2}[u(\theta) - \ve]}
\nonumber
\end{align}
For consistency, we use the same interaction energy $u(\theta) \eta(i)$ in \eqref{act} as in \eqref{act1}. The nonequilibrium is parametrized by the parameter
$\ve>0$ which breaks the left/right symmetry.  For large $\ve$ a bit gets typically loaded $0\rightarrow 1$ from the right channel, and typically decharges (``fires'') via the left channel.  There it gets coupled with the angle-coordinate via the (average) rate $a_\ell$.  In that way, the initial (digitalized) signal gets transduced ``from the right'' to the more macroscopic variable $\theta$ ``to the left,'' see the representation in Fig.~\ref{fig2}.  

 For the average firing rate we choose $a_\ell(\theta) = a(1 + \lambda\,b\, \cos\theta)>0$ with reference frequency $a$, excess $b>0$ and small dimensionless coupling $\lambda\geq 0$ with $\lambda b<1$.    Again we think of $b=b(I)$ as a coding of stimulus $I$.  The equations \eqref{act}--\eqref{act1} define the active dynamics with $\eta_t(i)$ playing the role of dichotomous noise as in the telegraph equation or for run-and-tumble particles; see \cite{nong}.\\
Assuming that the bit relaxes almost instantaneously on the time-scale of the angle-coordinate (large $a$), the dynamics in \eqref{act} becomes $\dot \theta_t = f(\theta_t);  f(\theta) := -U'(\theta)- Nu'(\theta) \rho_\theta(1)$ where $\rho_\theta(1)$ is the stationary probability that the bit is loaded ($\eta(i)=1$).
We easily deduce it from \eqref{act1},
\[
\rho_\theta(1) = \frac 1{1+  e^{u(\theta)}(1 + \lambda\,b\, \cos\theta)}
\]
in the limit of large $\ve$ \cite{nong}.  For $b=0$ the force $f(\theta)= -V'(\theta)$ is derivable from a potential $V$; there can be no angular current then. On the other hand, for large $b$, a loaded bit fires almost instantly along the left channel. The rotational part of the quasistatic force $f$ is its circle integral
\[
F(b) = -\oint \id \theta\,\frac{Nu'(\theta)} {1+  e^{u(\theta)}(1 + \lambda\,b\, \cos\theta)}
\]
which is essentially linear in $b$ for weak coupling $\lambda$,
\[
F(b) = K\, b, \qquad K :=\frac{\lambda N}{2} \oint \id \theta\,\sin\theta\,\tanh (u(\theta)/2)
\]
As a consequence, whenever $K\neq 0$ and for $b>b_0$, we have many choices of reference potential $U$ so that $f(\theta) > kb$ for some $k>0$ and there appears an angular current (i.e., perceived output) which is proportional to $b$, meaning that the perceived stimulus $I_p\propto b(I)$.
The $b_0$ corresponds to the just-noticeable-difference in Weber's law.   That is the analogue of \eqref{vel}.  The only remaining assumption to obtain Weber-Fechner behavior is again to assume that $b \propto \log I$ in a range of stimuli $I$.

\section{General argument}\label{nonr}
The wonderful universality of the Weber-Fechner laws begs for a principled answer as to why they hold.  As announced from the above models, there are two main ingredients.\\

One, to be discussed next under subsection \ref{kex}, is the understanding how a noticeable displacement is achieved from and is proportional to the change in reactivity.  The conceptual framework is found in response theory around nonequilibria.
Stimuli are understood as perturbing a stationary condition of an open possibly far-from-equilibrium system.  Even while fixing the thermodynamic (mostly electrochemical) force, kinetics may change.   We interpret the strength of the resulting displacement, when exceeding some threshold, to be the {\it perceived} stimulus.  In the previous section, in the passive model $I_p$ was the displacement in current; in the active model $I_p$ was the induced angular velocity.  The relation between perceived stimulus and actual stimulus, the central topic of the Weber-Fechner laws, thus gets formulated as a subject in response theory.\\

The second ingredient, to be discussed under subsection \ref{webr}
is the quantitative identification of excess reactivity with the logarithm of the actual stimulus. We argue there that the excess reactivity is proportional to the logarithm of the actual stimulus, giving two arguments --- one using biased diffusion as decision model and one based on the naturalness of logarithmic conversion from analogue to digital signaling. These two arguments do not share the rigor of the next section (first ingredient) and corrections such as in Stevens' law are perfectly compatible and possible. 

\subsection{Kinetic excess}\label{kex}
The models in the previous section are explicit because they are simple.  Yet they are on target for the dependence on kinetics.   At high driving ($\ve\gg 1$), the reactivities gain centerstage.   That dependence on kinetics (motility, trapping configurations,...) is a typical nonequilibrium feature, which is both source of difficulties and of richness, \cite{fren,springer,mar,maarten}. The main point is that the response formul{\ae} not only feature the dissipation caused by the stimulus, but also correlate with the excess dynamical activity, called the frenetic contribution; see \cite{resp,urna,fren,maarten}.  The usual McLennan-Zubarev ensembles describing close-to-equilibrium physics do not suffice for that purpose; far from nonequilibrium, displacements and differential susceptibilities are determined by (changes in) the time-symmetric dynamical activity.  That also makes the main point of the paper: the actual stimulus is encoded as excess dynamical activity in a nonequilibrium system, causing extra displacement which is read as the perceived stimulus.\\

The argument is presented with some technical details.  To be specific, and keeping the context of neuronal activity we consider a general Markov jump process wih states $x,y,\ldots$ and transition rates
\begin{equation}\label{rara}
k(x,y) = a(x,y) \,e^{\sigma(x,y)},\qquad a(x,y)=a(y,x),\;\; \sigma(x,y)=-\sigma(y,x)
\end{equation}
for the jump $x\rightarrow y$.  We call the symmetric prefactors $a(x,y)$ reactivities, obviously related to escape rates and reaction times.  In the exponential we place antisymmetric $\sigma(x,y)$ which quanitify the nonequilibrium aspect; they are not given from a difference $\sigma_\text{eq}(x,y) = V(x) - V(y)$ in some state function $V$.  The absence of such a potential $V$ means that the condition of detailed balance is violated.  We do not specify the origin of that breaking and many examples may be considered.  Yet, the breaking of detailed balance is crucial for the role of the reactivities.  The main question is now --- Suppose we have such a nonequilibrium system which is perturbed by changing its reaction rates:  how can that affect or generate currents?  \\

Consider an observable $F(\omega)$ which depends on the random trajectory $\omega = (x_s, 0\leq s\leq t)$ in time-window $[0,t]$.  Every such $\omega$ consists of jumps separated by quiescent periods.  We take the $F(\omega)$ to be time-extensive and antisymmetric under time-reversal as befits time-integrated particle or energy currents.  We call $F$ a displacement.  We look at expectations, taking the mean over many such processes \eqref{rara}.  Without loss of generality we suppose that $\langle F\rangle=0$ in the stationary nonequilibrium condition.  The question is to see how that expectation changes when we modify the reactivity, say $a(x,y) \rightarrow a(x,y)(1+ \lambda\,b(x,y))$, where the $b(x,y)=b(y,x)$ is seen as the perturbation.  The answer from nonequilibrium response theory \cite{resp} is that  we get a (new) expected displacement $\langle F\rangle_b$, to linear order in $\lambda$, equal to
\begin{equation}\label{respl}
\langle F\rangle_b = -\lambda\, \langle D(\omega)\,F(\omega)\rangle
\end{equation}
in terms of the unperturbed expectation $\langle \cdot\rangle$, correlating the discplacement with $D(\omega)$,  the frenesy given by
\begin{equation}\label{sumd}
D(\omega) = \int_0^t\id s \, \sum_ya(x_s,y)b(x_s,y)e^{\sigma(x_s,y)} - \sum_{s} b(x_{s^-},x_{s})
\end{equation}
where the last sum is over the jump times in $\omega$.  The point is already visible: the response is proportional to the excess or additional reactivity.  That would not be true in the case of detailed balance; then, around equilibrium $\langle \cdot\rangle =\langle \cdot\rangle_\text{eq}$ we have $\langle D(\omega)\,F(\omega)\rangle_\text{eq} =0$ by time-reversal invariance since $D$ is time-symmetric.  Secondly, the response relation \eqref{respl} decisively uses that the antisymmetric factors $\sigma(x,y)$ have not been modified.  We refer to \cite{resp,fren,springer} for more details.  Taking the situation far from equilibrium as parametrized by $\sigma(x,y) =\ve \,v(x,y)$ with $\ve\uparrow \infty$, we only retain in the first sum \eqref{sumd} the transitions $x\rightarrow y$ for which $\sigma(x,y) >0$.  Moreover, to allow for saturation of the mean displacement $\langle F\rangle_b$ as  $\ve\uparrow \infty$, we are ready to take $a(x,y)$ to depend on $\ve$ so that $a(x,y)\, \exp \ve v(x,y)\rightarrow a$ whenever $v(x,y)\geq 0$.  We conclude that the far-from-equilibrium response \eqref{respl} gives 
\begin{equation}\label{perc}
\langle F \rangle_b = -\lambda\, \langle B(\omega)\,F(\omega)\rangle
\end{equation}
where
\[
B(\omega) :=a\int_0^t\id s \, \sum_{y: \sigma(x_s,y)>0} b(x_s,y) - \sum_{s} b(x_s,x_{s^+})
\]
is  proportional to the excess reactivity.  In the simplest scenario, the reaction channels do not change under the applied stimulus $I$ and we simply put $b(x,y) = b(I) >0$ constant, yielding
\[
B(\omega) = \big[a \int_0^t n_s - N_t\big]\,b(I)
\] 
for $n_s$ the number of available $y$ where $\sigma(x_s,y)>0$ and $N_t$ the total number of jumps in $\omega$.  In all, we get
\begin{equation}\label{perc}
f_b= K\,b(I), \qquad K := \lambda\, \lim_t \frac 1{t}\langle [N_t-a \int_0^t n_s ]\,F(\omega)\rangle
\end{equation}
for the expected flux $f_b := \lim_t \langle F\rangle_b/t$  under a change in reactivity $b$.  The $f_b$ can be directly identified with the perceived stimulus $I_p$ or we can still use that $f_b$ to create an angular velocity which is then taken as $I_p$. For example in a dynamics for an angle $\theta \in [0,2\pi]$  of the form
\begin{equation}\label{cr}
\dot\theta = -U'(\theta) + f_b,\quad I_p =\left[ \oint \frac1{ -U'(\theta) + f_b} \id \theta\right]^{-1}\propto b(I)
\end{equation}
the $f_b$ functions as a force.
Clearly, the rotational force $f_b$ causes a current $I_p$ for $b$ large enough, to overcome a possible barrier imposed by potential $U$.  That defines the just-noticeable excess $b_0$, as we interpret the emerging current or velocity as the perceived stimulus.
The choice of dynamics \eqref{cr} is not the most general one, nor does it share neurobiological complexity.  Yet, details or variations on the same theme do not affect the general conclusions. 

\subsection{The logarithm}\label{webr}
Stimulus starts by currents being injected into the dendritic tree and being transformed into a train of spikes.  That response of a neuron to synaptic input is not the main subject of the present paper but must be considered to complete the argument.\\  
 The states $x$ in the Markov process \eqref{rara} represent a large number of chemomechanical configurations possibly involving many interacting components.
Reaction rates for neurons are indeed the result of multi-level activities as they involve stimulus processing, decision making, and response programming. It often involves cognitive aspects as well. Information flow within an organism is of course a complicated issue.  Nevertheless, gaining simplicity from our mathematical modeling, it is not unreasonable to locate the dependence of reaction times in the symmetric prefactor (reactivity) $a(x,y)$ of the transition rates.  How then does spiking or the firing rates depend on the magnitude of the stimulus?\\

Clearly, the question belongs to the field of neural coding, where one possible approach is called rate coding.  In many cases and over various regimes, as the intensity of a stimulus increases, the frequency of ``spike firing'' increases.
In other words, there is good experimental evidence that reaction times decrease with the magnitude of the stimulus; see e.g. \cite{neuro,nis}. Whether there is a logarithmic dependence is however a more subtle and detailed issue.   For the optimality of the logarithmic scale from the point of view of evolution of the cognitive apparatus, see e.g. \cite{port}.   It is in general agreed that  the  distribution  of  spike rates  within  any  neural  tissue  follows  a  lognormal  distribution, whether at rest or when stimulated; see \cite{schel}.  However, each individual neuron is rather stable in its firing rate, and variability  in  mean  spike  rate  is to be understood over  the  whole population.   Neurons spike with 5–10fold different mean rates which increases the dynamic range of a neuronal population.
 Therefore, logarithmic dependence of firing rates on stimulus strength should be seen in the ensemble sense.   
 We give two rather general theoretical arguments for the dependence $b(I) \sim \log I$ of the mean excess reactivity $b$ on the stimulus $I$.\\

We start with heuristics belonging to a diffusion model for reactivities.\\
A widely-used model for estimating reaction times is to have a decision variable $X_t$ to undergo biased diffusion in the interval $[-1,1]$; see for example Section 10 in \cite{rat}.
 The time to reach the upper boundary (if at all) gives the reaction time.  Here again we can use statistical mechanics as the question relates to first-passage problems.   More specifically we look at splitting probabilities representing reactivities and we know that those probabilities vary linearly in the bias.  See e.g.~Eq.~2.2.11 in  \cite{red} where the bias is represented by the initial position $X_0$ and the splitting probability varies linearly in the relative distance between $X_0$ and the boundary.  Similarly, when the bias is represented as the gradient $\nabla \mu$ of a chemical potential, the reaction time is linear in $\nabla \mu$.  In all events, the linearity implies that the question becomes how the bias (hence the excess reactivity $b(I)$) in that diffusion process should depend on the stimulus $I$.\\
 Continuing the diffusion analogy and language where the bias represents the gradient in chemical potential, we think of the stimulus $I$ as the extra density which causes the gradient. In other words, the stimulus $I$ corresponds to an extra density or pressure at one end of the gas tube, where the bias $b$ making the asymmetry is a gradient in chemical potential.  Similarly, the initial position $X_0$ represents the difference in chemical potential between the two ends of the interval causing a drift.  The conclusion mimics therefore the logaritmic dependence of chemical potential on density or pressure in a free gas. Driving in irreversible thermodynamics is achieved from gradients in chemical potential which vary logarithmically with the density (stimulus).\\

Secondly and as prime statistical argument, we consider the conversion of an analogue signal (stimulus) into a digital one (neuronal activity).  A stimulus of strength $I$ is causing neurons in a certain region to fire more often per unit time.  There is no change in nonequilibrium driving, that remains provided by biochemical engines such as the hydrolysis of ATP. Firing encodes the stimulus, and the stimulus selects the neuronal activity.  Clearly, storing information of a signal with strength $I$ requires $\log I$ binary signals, to be divided over neurons and their firing rate.  In that way, the spacetime activity is of (binary nature) of order $\log I$.  In the end and similar to the logarithm defining Boltzmann's entropy, $\log I$ is the number of bit-variables over spacetime for representing the stimulus $I$.  As the spatial region for neuronal action is limited, the storage happens extensively in the temporal domain, by increasing the firing rate in proportion with $\log I$.\\

\subsection{Limitations}\label{lims}
We summarize the main line of reasoning.  The Weber-Fechner laws can be stated in the framework of response theory around nonequilibrium. It is assumed that a stimulus $I$ only affects the frenetic component, i.e., causing an excess $b(I)$ in the time-symmetric dynamical activity or reactivities. Far-from-equilibrium response theory shows that the induced displacement or current is proportional to that excess frenesy $b(I)$.  We identify that displacement with the perceived stimulus $I_p$.  If the stimulus strength $I$ is converted into neuronal activity by shifting the mean reactivity by order $b(I) \sim \log I$, then the above suffices for arguing Fechner's law, and so follows Weber's law.\\

We mention some limitations of the above reasonings.  First, we have assumed thoughout that the coupling between the applied stimulus and the reaction rate is sufficiently small. That was evidenced from the small coupling $\lambda$ and the application of linear response.  That obviously also limits from above the value of $b$ and hence it bounds the allowed $\log I$ from above: we cannot expect to find Weber-Fechner behavior for very large values of $\log I$ (but $I$ itself can be relatively big). Secondly, we have not only assumed conditions which break the fluctuation-dissipaton relation (which is needed) but we have for simplicity taken far-from-equilibrium reference conditions.  That is probably not necessary but it will correct the strict linearity of the response $f_b$ in the excess reactivities $b$. In the same way, we assumed that the dynamics of the collective coordinate (the angle $\theta$) through which the stimulus becomes perceived, is not noisy.  That zero-temperaure condition is again not truly needed but corrections will of course occur, be it small ones.  Finally, while the upshot of much experimental work on reactivity in neuronal networks is compatible with the hypothesis that the stimulus causes an excess in reaction rate which scales logarithmically in the actual stimulus, $b \propto \log I$, there is also evidence of power law behavior.  Then, the function $b(I)$ scales with some power of $I$ in certain regimes, obviously and straightforwardly modifying the Weber-Fechner laws in the direction of e.g. Stevens' law, \cite{stevr,rev}.  Moreover, the increase in neuronal firing rate (e.g. by reducing the membrane time constant when modeled as capacitance) is obviously not unlimited; for very high synaptic input, firing rates tend to decrease again (e.g. by shunting membrane potential fluctuations).

\section{Conclusions}
  With the advent of nonequilibrium fluctuation theory, new avenues have been opened to solving problems and understanding phenomena of biophysics and psychophysics. Part of that is nonequilibrium response theory that we use in this paper as framework for discussing the Weber-Fechner laws. The central observation is that the dependence of reactivities on the actual stimulus decides the response when the system is far enough from equilibrium.  Weber-Fechner phenomenology then follows from two hypotheses (1) that the perceived stimulus is a displacement (or current) in response to changes in neuronal activity, (2) that mean excess in the addressed ensemble of neuronal reactivities scales with the logarithm of the actual stimulus.  Corrections to (2) such as a power-law dependence of reactivity on stimulus lead to Stevens' law.\\

\end{document}